\documentclass[12pt]{article}
\usepackage{fullpage}
\usepackage{amssymb} 
\usepackage[all,dvips]{xy} 
\usepackage{url}
\usepackage{pstricks}

\usepackage{theorem}
\newtheorem{theo}{Theorem}[section]
\theorembodyfont{\normalfont}
\newtheorem{defi}[theo]{Definition}
\newtheorem{exam}[theo]{Example}

\newcommand{\Tt}{\Theta} 
\renewcommand{\tt}{\theta} 
\newcommand{\Ss}{\Sigma} 
\renewcommand{\ss}{\sigma} 
\newcommand{\fr}[2]{{#2}\backslash{#1}} 
\newcommand{\To}{\Rightarrow} 

\newcommand{\catC}{\mathbf{C}}   
\newcommand{\catS}{\mathbf{S}} 
\newcommand{\catT}{\mathbf{T}} 
\newcommand{\Set}{\mathbf{Set}} 
 
\newcommand{\DiaLog}{\mathbf{DiaLog}} 
\newcommand{\skE}{\mathbf{E}}  
\newcommand{\ske}{\mathbf{e}}  

\newcommand{\Rea}{\mathit{Real}} 
\newcommand{\Mod}{\mathit{Mod}} 
\newcommand{\Elt}{\mathit{Elt}} 

\newcommand{\funY}{\mathcal{Y}} 
\newcommand{\Hh}{\mathcal{H}} 
\newcommand{\Cc}{\mathcal{C}} 

\newcommand{\eq}{\mathrm{eq}} 
\newcommand{\far}{\mathrm{far}} 
\newcommand{\dec}{\mathrm{dec}} 
\newcommand{\near}{\mathrm{near}} 
\newcommand{\id}{\mathrm{id}} 
\newcommand{\pr}{\mathrm{pr}} 

\newcommand{\tti}{\mathtt{i}}

\newcommand{\ttiz}{\mathtt{i0}}
\newcommand{\ttpm}{\mathtt{c}}
\newcommand{\Type}{\mathtt{Type}}
\newcommand{\Term}{\mathtt{Term}}
\newcommand{\dom}{\mathtt{dom}}
\newcommand{\codom}{\mathtt{codom}}
\newcommand{\selid}{\mathtt{selid}}
\newcommand{\Selid}{\mathtt{Selid}}
\newcommand{\Cons}{\mathtt{Cons}}
\newcommand{\fst}{\mathtt{fst}}
\newcommand{\snd}{\mathtt{snd}}
\newcommand{\comp}{\mathtt{comp}}
\newcommand{\Comp}{\mathtt{Comp}}

\newcommand{\bS}{\mathbb{S}} 

\def\ltimesseq{\mathrel{\vcenter{\hbox{\psset{unit=0.3mm,linewidth=0.2}\begin{pspicture}(0,-4)(7,4)\pspolygon[fillstyle=solid,fillcolor=black](0,-3)(0,3)(3,0)\psline(6,3)(0,-3)(0,3)(6,-3)\end{pspicture}}}}}

\title{How to combine diagrammatic logics}  
\author{Dominique Duval \\ \small{
LJK, Universit\'e de Grenoble, France, \url{Dominique.Duval@imag.fr}} 
}
\date{November 19., 2009}

\begin{document}
\maketitle

\textit{
This paper is a submission to the contest: \textrm{How to combine logics?}
at the World Congress and School on Universal Logic III, 2010.
In this version~2, some typos have been corrected.}


\section{Introduction}

We claim that combining ``things'', whatever these things are, 
is made easier if these things can be seen as the objects of a category.
Then things are related by morphisms,
and the categorical machinery provides various tools for combining things,
typically limits and colimits, as well as other constructions. 

What should be required from an object in a category to call it a logic? 
The analysis of various kinds of logic brings in light similar issues:
which are the sentences of interest? 
what is the meaning of each sentence? 
how can a sentence be infered from another one?
According to this analysis, 
each logic should have a syntax, a notion of model, a proof system, 
and the morphisms should, in some sense, preserve them. 
In section~\ref{sec:dialog} we define the 
\emph{diagrammatic logics}, which satisfy the required properties.
Then in section~\ref{sec:combine} 
the category of diagrammatic logics follows, 
so that categorical constructions can be used for combining diagrammatic logics.
As an example, a combination of logics using an opfibration is presented,
in order to study computational side-effects
due to the evolution of the state 
during the execution of an imperative program.

This work stems from the study of the semantics 
of computational effects in programming languages.
It has been influenced by 
the theory of sketches \cite{Eh68,Lellahi89}
and the sketch entailments \cite{Makkai97}, 
by categorical logic \cite{Law63}, 
the theory of institutions \cite{GB84}, 
the use of monads in computer science \cite{Moggi89},
and many other papers and people. 
Diagrammatic specifications appeared in \cite{Du03}.

\section{Diagrammatic logics}
\label{sec:dialog}

\subsection{Theories and specifications}

The word ``theory'' is widely used in logic, with various meanings.
In this paper, a \emph{theory} for a given logic is a class of sentences
which is saturated: it is a class of theorems from which no new theorem 
can be derived inside the given logic.
For instance, the theory of groups in equational logic is made of all the theorems
which can be proved about groups, starting from the usual equational axioms for groups
and using the congruence properties of equality.
There are morphisms between theories: 
for instance, there is an inclusion of the theory of monoids in the theory of groups. 
A class of axioms generating a theory is often called a ``presentation'' 
or a ``specification'' of this theory (according whether this notion is used by 
a logician or a computer scientist), in this paper it is called a \emph{specification}.
So, in a given logic, every specification generates a theory,
while every theory may be seen as a (huge) specification.
In categorical terms, this means that the category $\catT$ of theories
is a \emph{reflective subcategory} 
of the category $\catS$ of specifications.
Or equivalently, this means that 
there is an \emph{adjunction} between $\catS$ and $\catT$, 
such that the right adjoint is full and faithful.
The right adjoint $R\colon \catT\to\catS$ allows to consider a theory as a specification,
the left adjoint $L\colon \catS\to\catT$ generates a theory from a specification,
and the fact that $R$ is full and faithful means that each theory is saturated.
So, let us define an  \emph{``abstract diagrammatic logic''}
as a functor $L$ with a adjoint $R$ which is full and faithful
(the actual definition of a diagrammatic logic, with a syntax, is definition~\ref{defi:dialog}).
 $$ \xymatrix@C=4pc{ \catS \ar[r]^{L}_{\bot} & \catT \ar@/^2ex/[l]^{R} } $$

For instance, the equational logic is obtained 
by defining 
on one side the equational theories are the categories with chosen finite products 
and on the other side the equational specifications either 
as the signatures with equations or (equivalently) as the finite product sketches.

Categorical logic usually focuses on theories rather than specifications,
and requires theories to be some kind of categories:
theories for equational logic are categories with finite products, 
theories for simply-typed lambda-calculus are cartesian closed categories, and so on. 
Indeed, specifications do not matter as long as one focuses on 
abstract properties of models, 
but they are useful for building explicitly such models (section~\ref{subsec:model}). 
More importantly, the mere notion of proof does rely on specifications: 
a specification is a class of axioms for a given theory, 
and a proof is a way to add a theorem to this class of axioms, 
which means that a proof is an enrichment of the given specification 
following the rules of the logic (section~\ref{subsec:proof}). 

\subsection{Models}
\label{subsec:model}

Now, let $L\colon \catS\to\catT$ be some fixed diagrammatic logic,
with right adjoint $R$.
A \emph{(strict) model} $M$ of a specification $\Ss$ in a theory $\Tt$ is 
a morphism of theories $M\colon L\Ss \to \Tt$.
Equivalently, thanks to the adjunction, a model $M$ of $\Ss$ in $\Tt$ is 
a morphism of specifications $M\colon \Ss \to R\Tt$. 
The abstract properties of models can be derived from 
their definition as morphisms of theories,
while the explicit construction of models 
makes use of their definition as morphisms of specifications.

For example in equational logic, 
let $\Tt$ be the category of sets considered as an equational theory:
then we recover the usual notion of model of an equational specification. 
For this logic, in addition, 
the natural transformations provide an interesting
notion of morphisms between models. 

\subsection{Instances}
\label{subsec:instance}

There are not ``enough'' morphisms of specifications, in the following sense: 
given specifications $\Ss$ and $\Ss_1$, 
usually only few morphisms of theories $\tt\colon L\Ss\to L\Ss_1$
are of the form $\tt=L\ss$ for a morphism of specifications $\ss\colon \Ss\to \Ss_1$.
In order to get  ``enough'' morphisms between specifications,
we introduce instances as generalizations of morphisms.

\begin{defi}
\label{defi:instance}
An \emph{entailment} for a diagrammatic logic $L\colon \catS\to\catT$
is a morphism $\tau\colon \Ss\to\Ss'$ in $\catS$ such that $L\tau$ is invertible in $\catT$.
This is denoted $\xymatrix{\Ss\ar[r]^{\tau} & \Ss' \ar@<1ex>@{-->}[l] \\ }$. 
An \emph{instance} $\kappa$ of a specification $\Ss$ in a specification $\Ss_1$
is a cospan in $\catS$ made of 
a morphism $\ss$ and an entailment~$\tau$
 $$ \xymatrix@C=4pc{
  \Ss \ar[r]^{\ss} &  
  \Ss'_1 \ar@{-->}@<-1ex>[r] & 
  \Ss_1 \ar[l]_{\tau}   \\  
  }$$ 
\end{defi}

Two specifications which are related by entailments are equivalent,
in the sense that they generate isomorphic theories.
An instance $\kappa=(\ss,\tau)$ is also called a \emph{fraction} 
with \emph{numerator} $\ss$ and 
\emph{denominator} $\tau$, it is denoted $\kappa=\fr{\ss}{\tau}\colon \Ss\to\Ss_1$.
Then $L\kappa$ is defined as $L\kappa=(L\tau)^{-1}\circ L\ss\colon L\Ss\to L\Ss_1$.
According to \cite{GZ67}, since the right adjoint $R$ is full and faithful, 
every morphism of theories $\tt\colon L\Ss\to L\Ss_1$ is of the form $\tt=L\kappa$ 
for an instance $\kappa$ of $\Ss$ in $\Ss_1$.
So, up to equivalence, the category $\catT$ is the category of fractions 
of $\catS$ with denominators the entailments. 

\subsection{Syntax}
\label{subsec:syntax}

The syntax for writing down the axioms and the inference rules of a given logic 
is provided by \emph{limit sketches} \cite{Eh68,BW99}.
A limit sketch $\skE$ is a presentation for a category with prescribed limits. 
Roughly speaking, a limit sketch $\skE$ is a graph with \emph{potential} 
composition, identities and limits, which means that the usual notations for 
composition, identities and limits can be used, 
for example there may be an arrow $g\circ f\colon X\to Z$ 
when $f\colon X\to Y$ and $g\colon Y\to Z$ are consecutive arrows, 
there may be an arrow $\id_X\colon X\to X$ when $X$ is a point,
there may be a point $Y_1\times Y_2$ with arrows 
$\pr_1\colon Y_1\times Y_2 \to Y_1$ and $\pr_2\colon Y_1\times Y_2 \to Y_2$ 
when $Y_1$ and $Y_2$ are points, and similarly for other limits.
But such arrows may as well not exist, and when they exist they do not 
have to satisfy the usual properties, 
for example the composition does not have to be associative.
A limit sketch $\skE$ generates a category $P(\skE)$ 
where each potential feature becomes a real one:
an arrow $g\circ f$ in $\skE$ becomes the real composition of $f$ and $g$ in $P(\skE)$,
an arrow $\id_X$ in $\skE$ becomes the real identity of $X$ in $P(\skE)$,
a point $Y_1\times Y_2$ with the arrows $\pr_1$ and $\pr_2$ in $\skE$ 
becomes the real product of $Y_1$ and $Y_2$ with its projections in $P(\skE)$,
and similarly for other limits.

A \emph{(set-valued) realization} (or \emph{model}) of 
a limit sketch $\skE$ interprets each 
point of $\skE$ as a set and each arrow as a function,
in such a way that potential features become real ones. 
The realizations of $\skE$ form a category $\Rea(\skE)$.
The \emph{Yoneda contravariant functor} $\funY_{\skE}$ 
is such that for each point $X$ of $\skE$ the realization
$\funY_{\skE}(X)$ is made of the arrows in $P(\skE)$ with source $X$.
This allows to identify the opposite of $\skE$ to a subcategory 
$\funY_{\skE}(\skE)$ (or simply $\funY(\skE)$) of $\Rea(\skE)$. 
This subcategory is \emph{dense} in $\Rea(\skE)$,
which means, for short, that 
every realization of $\skE$ is a colimit of realizations in $\funY(\skE)$.

Every morphism of limit sketches $\ske\colon \skE_S\to\skE_T$
defines an adjunction between the categories of realizations,
where the right adjoint $R_{\ske}\colon \Rea(\skE_T)\to\Rea(\skE_S)$ 
is the precomposition with $\ske$,
which may be called the \emph{forgetful} functor with respect to $\ske$.
Then the left adjoint $L_{\ske}\colon \Rea(\skE_S)\to\Rea(\skE_T)$ extends $\ske$,
contravariantly,
it may be called the \emph{freely generating} functor with respect to $\ske$.
A category is \emph{locally presentable} when it is equivalent 
to $\Rea(\skE)$ for some limit sketch $\skE$, 
and we say that a  functor is \emph{locally presentable}
when it is (up to equivalence) the left adjoint functor $L_{\ske}$
for some morphism of limit sketches $\ske$.
More precisely, we assume that 
a locally presentable category comes with a presentation $\skE$,
and a locally presentable functor with a presentation $\ske$.
Now we can define diagrammatic logics.

\begin{defi}
\label{defi:dialog}
A \emph{diagrammatic logic} is a locally presentable functor $L$
such that its right adjoint $R$ is full and faithful.
An \emph{inference system} for $L$
is a morphism of limit sketches $\ske\colon \skE_S\to\skE_T$ 
such that $L=L_{\ske}$. 
\end{defi}
Thanks to the Yoneda contravariant functor, 
the morphism $\ske$ and the functor $L_{\ske}$ have similar properties.
In particular, $\ske$ can be chosen so as to 
consist of adding inverse arrows for some arrows in $\skE_S$.
In addition, since every locally presentable category has colimits, 
the composition of instances in $\catS$ can be performed using pushouts. 

\subsection{Rules and proofs}
\label{subsec:proof}

Given a  rule $\frac{\Hh_1 \;\dots\; \Hh_n}{\Cc}$ in a given logic,
the hypotheses $\Hh_1,\dots,\Hh_n$ as well as the conclusion $\Cc$
may be seen as specifications. 
The hypotheses may be amalgamated, as a unique hypothesis $\Hh$
made of the colimit of the $\Hh_i$'s
(it is not a sum usually: the relations between the $\Hh_i$'s
are expressed in the sharing of names).
Similarly, let $\Hh'$ be the colimit of $\Hh$ and $\Cc$.
So, without loss of generality, let us consider a rule $\frac{\Hh}{\Cc}$ in a given logic,
where the hypothesis $\Hh$ and the conclusion $\Cc$ are specifications 
with colimit $\Hh'$.
The meaning of this rule is that each instance of $\Hh$
in a theory $\Tt$ can be uniquely extended as an instance of $\Hh'$ in $\Tt$,
which means that the morphism $\Hh\to\Hh'$ is an entailment, 
then clearly this yields an instance of $\Cc$ in $\Tt$. 
So, basically, a rule is a fraction, i.e., it is an instance of $\Cc$ in $\Hh$. 
 $$ \xymatrix@C=4pc{
  \Hh \ar[r] &  
  \Hh' \ar@{-->}@<1ex>[l] & 
  \Cc  \ar[l]  \\  
  }$$ 
However, there is a distinction between an elementary rule and a derived rule, 
which will be called respectiveley a rule (or an inference rule) and a proof.

\begin{defi}
\label{defi:rule}
An \emph{inference rule} $\rho$ with \emph{hypothesis} $\Hh$ 
and \emph{conclusion} $\Cc$
is an instance $\rho=\fr{\ss}{\tau}\colon \Cc\to\Hh$ of the conclusion in the hypothesis, 
where $\ss$ and $\tau$ are in $\funY(\skE)$.
Given an inference rule $\rho=\fr{\ss}{\tau}\colon \Cc\to\Hh$ 
and an instance $\kappa\colon \Hh \to \Ss$ 
of the hypothesis $\Hh$ in a specification $\Ss$,
the corresponding \emph{inference step} 
provides the instance $\kappa\circ\rho\colon \Cc \to \Ss$ 
of the conclusion $\Cc$ in $\Ss$.
The commutative triangle in the category of fractions (on the left)
is computed thanks to the commutative diagram in the category 
of specifications (on the right) where the square is a pushout.
  $$ 
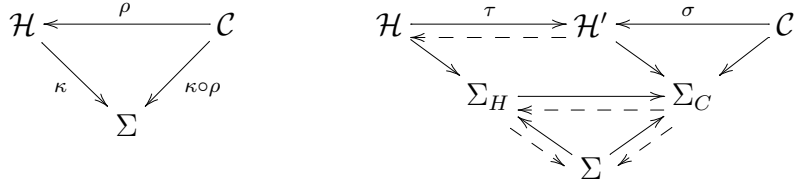
\xymatrix@C=2pc{
  \Hh \ar[dr]_{\kappa} & & \Cc \ar[ll]_{\rho} \ar[dl]^{\kappa\circ\rho} \\ & \Ss & \\ 
  } \qquad \qquad  
  \xymatrix@C=1.5pc@R=1pc{
  \Hh \ar[dr] \ar[rr]^{\tau} && 
    \Hh' \ar[dr] \ar@{-->}@<1ex>[ll] && 
    \Cc \ar[ll]_{\ss} \ar[dl] \\ 
  & \Ss_H \ar@{-->}@<-1ex>[rd] \ar[rr] & & 
   \Ss_C \ar@{-->}@<1ex>[ld] \ar@{-->}@<1ex>[ll] & \\ 
  & & \Ss \ar[lu] \ar[ru] & & \\ 
  }$$
A \emph{proof} (or \emph{derivation}, or \emph{derived rule}) 
is the description of a morphism of theories as an instance
composed from inference rules.
\end{defi}

It may be noted that the hypothesis is on the denominator side of the rule
and the conclusion on the numerator side,
in contrast with the usual notation $\frac{\Hh}{\Cc}$ 
which looks like a fraction with the hypothesis as numerator 
and the conclusion as denominator.

\begin{exam}
Let us consider the \emph{modus ponens} rule 
  $$ \frac{A \qquad A\To B}{B} $$
from the diagrammatic point of view. 
In a logic with this rule, 
a specification $\Ss$ is made of (at least) a set $F$ called the set of \emph{formulas},
a subset $P$ of $F$ called the subset of \emph{provable formulas},
and a binary operation $\To$ on formulas;
this will be denoted simply as $\Ss=(F,P)$ or even $\Ss=P$.
The modus ponens rule is obtained, essentially, 
by inverting one morphism of specifications. Let 
$$\Hh\!=\!(\{A,B,A\To B\},\{A,A\To B\}),\,
\Hh'\!=\!(\{A,B,A\To B\},\{A,A\To B,B\}),\, \Cc\!=\!(\{C\},\{C\})$$
where the names $A$, $B$ and $C$ stand for arbitrary formulas.
Let $\tau\colon \Hh\to\Hh'$ be the inclusion 
and $\ss\colon \Cc\to\Hh'$ the morphism which maps $C$ to $A\To B$. 
A theory is a specification where the modus ponens rule is satisfied, 
which means that $L\tau$ is an isomorphism,
i.e., that $\tau$ is an entailment. 
So, the diagrammatic version of the modus ponens rule is the fraction
 $$ \xymatrix@C=4pc{
  \Hh=\{A,A\To B\} \ar[r]^{\subseteq} &  
  \Hh'=\{A,A\To B,B\}  \ar@{-->}@<1ex>[l] & 
  \Cc=\{C\} \ar[l]_{\qquad\qquad C\mapsto A\To B}  \\  
  }$$ 
\end{exam}

\begin{exam}
An inference system $\ske_{\eq}\colon \skE_{\eq,S}\to\skE_{\eq,T}$ 
for equational logic is described in 
\\ \cite{DD09}. Focusing on unary operations,
for dealing with composition and identities the limit sketch $\skE_{\eq,S}$ contains
the following subsketch
$$
 
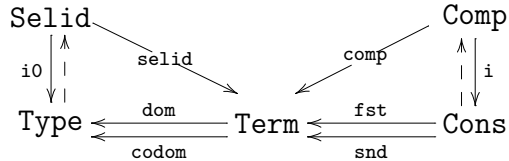
\xymatrix@C=4pc{
   \Selid \ar[d]_{\ttiz} \ar@<1ex>[rd]|{\selid} & &
       \Comp \ar[d]^{\tti} \ar@<-1ex>[dl]|{\comp} \\
   \Type \ar@<-1ex>@{-->}[u] & 
      \Term \ar@<1ex>[l]^{\codom} \ar[l]_{\dom} &
      \Cons \ar@<1ex>[l]^{\snd} \ar[l]_{\fst} \ar@<1ex>@{-->}[u]  \\
    }
$$
with the suitable potential limits, so that 
the image by Yoneda of this part of $\skE_{\eq,S}$
is the following diagram of equational specifications: 
$$
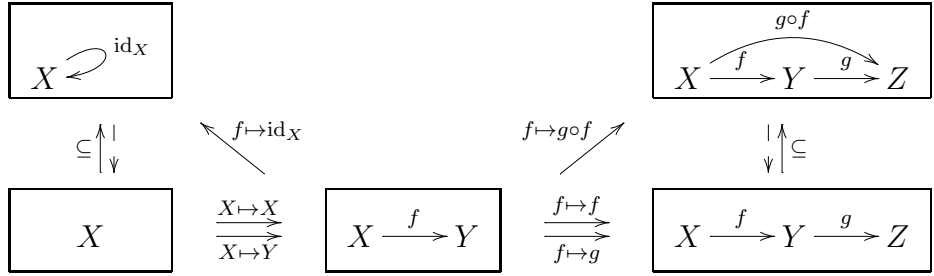
\begin{array}{|c|c|c|c|c|}
\cline{1-1} \cline{5-5} 
\xymatrix{X\ar@(ru,r)^{\id_X} } &
\multicolumn{3}{|c|}{} &
\xymatrix{X\ar[r]^{f} \ar@/^3ex/[rr]^{g\circ f} & Y\ar[r]^{g} & Z} \\
\cline{1-1} \cline{5-5} 
\multicolumn{1}{c}{\xymatrix@R=1.5pc{ \ar@<1ex>@{-->}[d] \\ \ar[u]^{\subseteq} \\}} &
\multicolumn{1}{c}{\xymatrix@R=1.5pc{ & \\ & \ar[ul]_{f\mapsto\id_X} \\}} & 
\multicolumn{1}{c}{} &
\multicolumn{1}{c}{\xymatrix@R=1.5pc{ & \\ \ar[ur]^{f\mapsto g\circ f}& \\}} &
\multicolumn{1}{c}{\xymatrix@R=1.5pc{ \ar@<-1ex>@{-->}[d] \\ \ar[u]_{\subseteq} \\}} \\ 
\cline{1-1} \cline{3-3} \cline{5-5} 
\xymatrix{X} &
\multicolumn{1}{|c|}{\xymatrix{ \ar@<1ex>[r]^{X\mapsto X} \ar[r]_{X\mapsto Y} & \\}} &
\xymatrix{X\ar[r]^{f} & Y} &
\multicolumn{1}{|c|}{\xymatrix{ \ar@<1ex>[r]^{f\mapsto f} \ar[r]_{f\mapsto g} & \\}} &
\xymatrix{X\ar[r]^{f} & Y\ar[r]^{g} & Z} \\
\cline{1-1} \cline{3-3} \cline{5-5} 
\end{array}
$$
The arrows $\tti$ and $\ttiz$ are entailments:
in the limit sketch $\skE_{\eq,T}$ there is an inverse for each of them.
This forms the rules 
  $$ \frac{f\colon X\to Y \quad g\colon Y\to Z}{g\circ f\colon X\to Y} \quad 
  \mbox{ and }\quad \frac{X}{\id_X\colon X\to X} $$
which means that in a theory 
every pair of consecutive terms can be composed
and every type has an identity.
The morphism $\ske$ is the inclusion $\skE_{\eq,S}\to\skE_{\eq,T}$.
\end{exam}

\section{Combining diagrammatic logics}
\label{sec:combine}

\subsection{Morphisms of logics}

\begin{defi}
\label{defi:morphism}
Let us consider two diagrammatic logics $L_1$ and $L_2$. 
A \emph{morphism} of diagrammatic logics $F\colon L_1\to L_2$ 
is a pair of locally presentable functors 
$(F_S,F_T)$ together with a natural isomorphism $F_T\circ L_1 \cong L_2\circ F_S$.
In practice, it is defined from inference systems 
$\ske_1\colon \skE_{S,1}\to\skE_{T,1}$ for $L_1$ 
and $\ske_2\colon \skE_{S,2}\to\skE_{T,2}$ for $L_2$ 
and from a pair of morphisms of 
limit sketches $\ske_S\colon \skE_{S,1}\to\skE_{S,2}$
and $\ske_T\colon \skE_{T,1}\to\skE_{T,2}$ which form a commutative diagram
  $$ \xymatrix@C=4pc@R=1.5pc{
  \skE_{S,1} \ar[r]^{\ske_1} \ar[d]_{\ske_S} & \skE_{T,1} \ar[d]^{\ske_T} \\
  \skE_{S,2} \ar[r]^{\ske_2} & \skE_{T,2} \\
  }$$
This defines the \emph{category of diagrammatic logics}, $\DiaLog$.
\end{defi}

Clearly a morphism of diagrammatic logics preserves the syntax
and the entailments, hence the rules and the proofs.
It does also behave well on models, thanks to adjunction:
let $U_S$ and $U_T$ denote the right adjoints of $F_S$ and $F_T$
respectively, then for each specification $\Ss_1$ for $L_1$
and each theory $\Tt_2$ for $L_2$, there is an isomorphism:
  \begin{equation}
  \label{eq:mod}
  \Mod_{L_1}(\Ss_1,U_T\Tt_2) \cong \Mod_{L_2}(F_S\Ss_1,\Tt_2) 
  \end{equation}

Now, combining diagrammatic logics can be performed by 
using categorical constructions in the category $\DiaLog$. 
These constructions may for example involve limits and colimits.
We now focus on opfibrations, 
because they can be used for dealing with computational effects,
which is our motivation for this work. 
Let us consider a typical computational effect: 
the evolution of the state during
the execution of a program written in an imperative language.

\subsection{Imperative programming} 
\label{subsec:state}

The state of the memory never appears explicitly in an imperative program,
although most parts of the program are written in view of modifying it.
In order to analyze this situation, 
we make a clear distinction between the commands as they appear
in the grammar of the language and the way they should be understood:
typically, 
in the grammar an assignment $X:=e$ has two arguments (a variable and an expression)
and does not return any value,
while it should be understood as a function with three arguments 
(a variable, an expression and a state) which returns a state.
Let $V,E,S$ stand for the sets of variables, expressions and states,
respectively, and $U$ for a singleton, then 
on one side $(:=)\colon V\times E \to U$ and 
on the other side $(:=)\colon S\times V\times E \to S$.
The fact of building a term $f\colon S\times X\to S\times Y$ 
from any term $f\colon X\to Y$ in a coherent way can be seen as a morphism of logics,
from the equational logic $L_{\eq}$ to the \emph{pointed} equational logic $L^*_{\eq}$,
made of the equational logic with a distinguished sort $S$. 

An operation like the assignment is called a \emph{modifier}, 
but there are also \emph{pure} operations that neither use nor modify the state,
like the arithmetic operations between the numerical constants.
Pure operations can be seen as special kinds of modifiers,
however the distinction between modifiers and pure operations 
is fundamental for dealing properly with the evolution of the state.
For example when the monad $(S\times-)^S$ on $\Set$
is used for this purpose, 
this distinction is provided by the inclusion of $\Set$ (for the pure operations)
in the Kleisli category of the monad  (for the modifers) \cite{Moggi89}.
This distinction can also be provided by indexing, or \emph{decorating},
each operation, with a keyword $p$ if it is pure or $m$
if it is a modifier. 
Then the decoration propagates to terms (pieces of programs)
in the obvious way: 
a term is pure if it is made only of variables and pure operations, 
otherwise it is a modifier. 
Because of the decorations, this logic is not the equational logic any more,
but a new logic called the \emph{decorated} equational logic $L_{\dec}$.
For example the basic part of the sketch for equational specifications (on the left)
is modified as follows (on the right),
where $\ttpm$ stands for the conversion of pure terms into modifiers.
$$
 
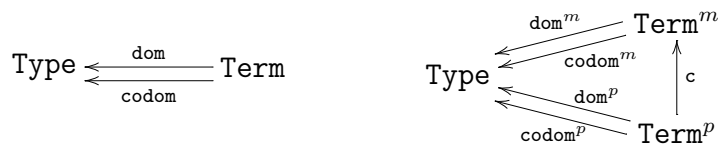
\xymatrix@C=4pc@R=0.5pc{
   \\  \Type  & 
      \Term \ar@<1ex>[l]^{\codom} \ar[l]_{\dom} \\ \\ 
    } \qquad\qquad 
 \xymatrix@C=4pc@R=0.5pc{
     & \Term^m \ar[ld]^(.4){\codom^m} \ar@<-1ex>[ld]_{\dom^m} \\
   \Type & \\
     & \Term^p \ar@<1ex>[lu]^{\codom^p} \ar[lu]_(.4){\dom^p} \ar[uu]_{\ttpm} \\
    } 
$$

\subsection{Zooms} 
\label{subsec:zoom}

In order to deal with the evolution of the state in section~\ref{subsec:state},
we used three different logics: 
the equational logic $L_{\eq}$ where the state is hidden,
the pointed equational logic $L^*_{\eq}$ for showing explicitly the state,
and the decorated equational logic $L_{\dec}$ 
as a kind of equational logic with additional information. 
These logics are related by morphisms, namely there is a span in the category $\DiaLog$ 
  $$ \xymatrix{
  & L_{\dec} \ar[dl]_{F_{\far}} \ar[dr]^{F_{\near}} & \\ 
  L_{\eq} && L^*_{\eq} \\
  }$$
The morphism $F_{\far}\colon L_{\dec}\to L_{\eq}$ simply forgets the decorations:
a modifier $f^m\colon X\to Y$ is mapped to $f\colon X \to Y$,
as well as a pure operation $f^p\colon X\to Y$.
This is the \emph{far} view, where the distinction between pure operations and
modifiers is blurred. 
The morphism $F_{\near}\colon L_{\dec}\to L^*_{\eq}$ 
provides the meaning of the decorations: 
a modifier $f^m\colon X\to Y$ becomes $f\colon S\times X \to S\times Y$
while a pure operation $f^p\colon X\to Y$ remains $f\colon X \to Y$.
This is the \emph{near} view, where the distinction between pure operations and
modifiers is explicited. 
This span should be read from left to right:
an equational specification $\Ss_{\eq}$ 
is derived from the grammar of the language,
then it is decorated as $\Ss_{\dec}$ such that $F_{\far}\Ss_{\dec}=\Ss_{\eq}$, 
and finally the meaning of the decorations is provided
by $\Ss_{\eq}^*=F_{\near}\Ss_{\dec}$. 
This span is called a \emph{zoom},
because it goes from the far view to the near view.

The semantics of the programs is given by a model of $\Ss_{\eq}^*$
in the pointed equational logic, with values in the theory $\Set_{\bS}$ made of 
the equational theory of sets together with some fixed set of states $\bS$
for interpreting the sort $S$.
Thanks to property~(\ref{eq:mod}), equivalently 
the semantics of the programs is given by a model of $\Ss_{\dec}$ 
in the decorated logic, with values in the decorated theory $U_{\near}\Set_{\bS}$,
where $U_{\near}$ is the right adjoint to $F_{\near}$.
But the semantics cannot be defined as a model of $\Ss_{\eq}$, 
which clearly does not bear enough details for providing a good semantics. 

The construction of $L_{\dec}$ and $F_{\far}$ from $L_{\eq}$ 
is a kind of \emph{opfibration}. 
Usually, given a functor $P\colon \catC\to\Set$, the 
\emph{category of elements} of $P$ is the category $\Elt(P)$ 
made of an object $X^x$ for each object $X$ in $\catC$ and each 
element $x\in P(X)$,
and with a morphism $f^x\colon (X,x)\to(Y,y)$ 
for each morphism $f\colon X\to Y$ in $\catC$ 
and each element $x\in P(X)$, where $y=Pf(x)$.
The functor $\varphi\colon \Elt(P) \to \catC$ which maps
$X^x$ to $X$ and $f^x$ to $f$ is called an \emph{opfibration}, 
it can be viewed as a $\catC$-indexed family of sets. 
This construction,  called the \emph{Grothendieck construction}, 
can be generalized in several ways. 
For our purpose, it can be generalized to limit sketches and to logics,
providing a systematic way for building new logics from well-known ones.
Given a logic $L_0$ and a theory $\Tt_0$ for this logic, 
the \emph{logic of elements} of $\Tt_0$ is a logic $L_1$ 
which can be seen as a variant of $L_0$ where every feature is 
indexed, or \emph{decorated}, by some feature in $\Tt_0$.
There is a morphism of logics $\varphi\colon L_1\to L_0$ 
which forgets the decorations.

For example, let $\Tt_0$ be an equational theory 
generated by a unique sort $D$ and two operations $p,m\colon D\to D$
such that $p\circ p=p$ and $p\circ m=m\circ p=m\circ m=m$.
Then the logic of elements of $\Tt_0$ looks like our decorated logic.
Actually, in order to get the conversion from pure terms to modifiers
in this way, we have to extend $\Tt_0$ as a theory which is not set-valued.
This does fit easily in our framework, because the realizations of
a sketch need not be set-valued, 
but this is beyond the scope of this paper. 

\subsection{Sequential products}

The zooming approach, as described in section~\ref{subsec:zoom},
was used in \cite{DDR09} 
for dealing with the issue of the order of evaluation of the arguments
of a binary function (or more generally a $n$-ary function with $n\geq2$).
When there is no side effect, a term like $g(a_1,a_2)$ can be seen as the composition of 
the function $g$ with the pair $(a_1,a_2)$,
which is formed from $a_1$ and $a_2$ using a cartesian product.
For simplicity of the presentation, let 
$a_i=f_i(x_i)$ for some function $f_i\colon X_i\to Y_i$, for $i=1,2$,
with $g\colon Y_1\times Y_2\to Z$.
Then $(a_1,a_2)=(f_1\times f_2)(x_1,x_2)$ and we focus on $f_1\times f_2$.
A cartesian product is a categorical product in the category of sets. 
The binary product on a category $C$ is 
such that for all $f_1\colon X_1\to Y_1$ and $f_2\colon X_2\to Y_2$
there is a unique $f_1\times f_2\colon X_1\times X_2 \to Y_1\times Y_2$ 
characterized by the following diagram, 
where the vertical morphisms are the projections
  $$
  \xymatrix@C=4pc{
  X_1 \ar[r]^{f_1} \ar@{}[rd]|{=} & Y_1 \\
  X_1\times X_2 \ar[r]^{f_1\times f_2} \ar[u] \ar[d] & 
  Y_1\times Y_2 \ar[u] \ar[d] \\
  X_2 \ar[r]^{f_2} & Y_2 \ar@{}[lu]|{=} \\
  }$$
When there are side effects, the effect and the result 
of evaluating $g(a_1,a_2)$ may depend on the  
order of evaluation of the arguments $a_1$ and $a_2$. 
This cannot be formalized by a cartesian product, which is a symmetric construction.
Another construction is required, for formalizing the fact of 
evaluating first $a_1$ then $a_2$ (or the contrary). 
So, the main issue is about evaluating $a_1$ while keeping $a_2$ unchanged. 
For this purpose, in \cite{DDR09} we go further into the decoration of 
equational specifications:
the terms are decorated as pure or modifiers as in section~\ref{subsec:state},
and in addition the equations themselves are decorated, 
as ``true'' equations (with symbol $=$) or \emph{consistency} equations (with symbol $\sim$).
On pure terms, both are interpreted as equalities.
On modifiers,
an equation $f^m = g^m\colon X\to Y$ becomes, through $F_{\near}$, 
an equality $f=g\colon S\times X\to S\times Y$,
whereas a consistency equation $f^m\sim g^m\colon X\to Y$ becomes 
an equality only on values (which is much weaker) 
$\pr\circ f=\pr\circ g\colon S\times X\to Y$, 
where $\pr\colon S\times Y\to Y$ is the projection.
Then the \emph{right semi-pure} product $f_1\rtimes \id_{X_2}$ 
is characterized by the following diagram in the decorated logic,
where the projections are pure
  $$
  \xymatrix@C=4pc{
  X_1 \ar[r]^{f_1^m} \ar@{}[rd]|{=} & Y_1 \\
  X_1\times X_2 \ar[r]^{(f_1\rtimes \id_{X_2})^m} \ar[u] \ar[d] & 
  Y_1\times X_2 \ar[u] \ar[d] \\
  X_2 \ar[r]^{\id_{X_2}^p} & X_2 \ar@{}[lu]|{\sim} \\
  }$$
The image of this diagram by the morphism $F_{\far}$
is such that the image of $(f_1\rtimes \id_{X_2})^m$ is $f_1\times \id_{X_2}$,
which maps $(s,x_1,x_2)$ to $(s_1,y_1,x_2)$ where $(s_1,y_1)=f_1(s,x_1)$.
Finally, the composition of a right semi-pure product  with a left semi-pure product 
gives rise to the required \emph{left sequential product}
$f_1 \ltimesseq f_2 = (\id_1 \ltimes f_2) \circ (f_1 \rtimes \id_2)$
for ``first $f_1$ then $f_2$''.
In the pointed equational logic $L_{\eq}^*$:
  $$ (f_1 \ltimesseq f_2)(s,x_1,x_2) = (s_2,y_1,y_2) 
  \mbox{ where } (s_1,y_1)=f_1(s,x_1) 
  \mbox{ and }  (s_2,y_2)=f_2(s_1,x_2) $$
This is depicted below,
first in the decorated logic $L_{\dec}$,
then through $F_{\near}$ in the pointed equational logic $L_{\eq}^*$.
  $$
  \xymatrix@C=4pc{
  X_1 \ar[r]^{f_1^m} \ar@{}[rd]|{=} & 
    Y_1 \ar[r]^{\id_{Y_1}^p} & 
    Y_1 \ar@{}[ld]|{\sim} \\
  X_1\times X_2 \ar[r]^{(f_1\rtimes \id_{X_2})^m} \ar[u] \ar[d] & 
    Y_1\times X_2 \ar[r]^{(\id_{Y_1}\ltimes f_2)^m} \ar[u] \ar[d] &
    Y_1\times Y_2 \ar[u] \ar[d] \\
  X_2 \ar[r]^{\id_{X_2}^p} & 
    X_2 \ar[r]^{f_2^m} \ar@{}[ru]|{=} \ar@{}[lu]|{\sim} &
    Y_2 \\
  }$$
  $$
  \xymatrix@C=4pc{
  S\times X_1 \ar[r]^{f_1} \ar@{}[rd]|{=} & 
    S\times Y_1  & 
    Y_1 \ar[r]^{\id_{Y_1}} & 
    Y_1 \ar@{}[ld]|{=} \\
  S\times X_1\times X_2 \ar[r]^{f_1\times \id_{X_2}} \ar[u] \ar[d] & 
    S\times Y_1\times X_2 \ar[u] \ar[d] \ar@{=}[r] &
    S\times Y_1\times X_2 \ar[r]^{\id_{Y_1}\times f_2} \ar[u] \ar[d] &
    S\times Y_1\times Y_2 \ar[u] \ar[d] \\
  X_2 \ar[r]^{\id_{X_2}} & 
    X_2 \ar@{}[lu]|{=} &
    S\times X_2 \ar[r]^{f_2} \ar@{}[ru]|{=} &
    S\times Y_2 \\
  }$$

\section{Conclusion}

The framework of diagrammatic logics stems from issues about 
the semantics of computational effects.
It is our hope that it may prove helpful for 
combining logics in other situations.


\end{document}